\documentclass[]{cupconf}
\usepackage{graphicx}

\title[Elemental abundance trends in the disks]
      {Elemental abundance trends in the \\ metal-rich 
thin and thick disks
}
      
\author[S. Feltzing]{\\ Sofia Feltzing
} 
\affiliation{Lund Observatory, Box 43, SE-221 00 Lund, Sweden
 }   

\begin{document}
\maketitle

\begin{abstract} Thick disks are common in spiral and S0 galaxies and
  seem to be an inherent part of galaxy formation and evolution. Our
  own Milky Way is host to an old thick disk.  The stars associated
  with this disk are enhanced in the $\alpha$-elements as compared to
  similar stars present in the thin disk. The Milky Way thin disk also
  appears to be younger than the thick disk.  Elemental abundance
  trends in stellar samples associated with the thin and the thick
  disks in the Milky Way are reviewed. Special attention is paid to
  how such samples are selected. Our current understanding of the
  elemental abundances and ages in the Milky Way thick and thin disks
  are summarised and discussed. The need for differential studies is
  stressed.  Finally, formation scenarios for the thick disk are
  briefly discussed in the light of the current observational picture.
\end{abstract}

\firstsection
\section{Introduction}

Thick disks appear to be ubiquitous in spiral and S0 galaxies (e.g.
Schwarzkopf \& Dettmar 2000; Dalcanton \& Bernstein 2002; Davidge
2005; Mould 2005; Elmegreen \& Elmegreen 2006) and the Milky Way is no
different as it hosts a thick disk in addition to the thin disk
(Gilmore \& Reid 1983).  The Milky Way thick disk has a scale-height
of about 1 kpc, which is three times that of the thin disk.

Available instrumentation and telescopes limits us to study the most
nearby stars if we wish to derive their elemental abundances and study
the chemical history of the Milky Way. Several stellar populations
overlap in the solar neighbourhood. The major components are the thin
and thick disks and the halo. There are also a multitude of streams
and so called moving groups.  The thick disk lags behind the local
standard of rest by $\sim 46$ km~s$^{-1}$ and the different disks have
different velocity dispersions. Using kinematic information we are
thus able to distinguish the thin and the thick disk stars from each
other (at least on a statistical basis).

It has proved very fruitful to combine kinematical and chemical
information for stars to derive the history of the disks in the Milky
Way. Edvardsson et al. (1993) provided one of the very first studies
to fully exploit this technique. Subsequent studies, utilising the
combination of kinematic and elemental abundance information, have
shown e.g.  that all thick disk stars that have been studied to date
are older than the thin disk stars (e.g. Bensby et al.  2003; Fuhrmann
2004). It is also now well established that stars, in the solar
neighbourhood, with kinematics typical of the thick disk are, at a
given [Fe/H], enhanced in $\alpha$-elements as compared to the thin
disk stars (e.g.  Bensby et al. 2004; Fuhrmann 2004, and
Fig.\ref{fig.oxygen}).

\section{How to define a thick disk star -- selecting stars for spectroscopy}

Stars in the thick disk rotate more slowly, in the plane, around the
galactic centre than the thin disk stars. The thick disk stars also
move higher above the galactic plane than the thin disk stars. The
velocity dispersions in all three galactic velocities for the thick
disk are also larger than the equivalent dispersions for the thin disk. 

In the solar neighbourhood we see a mixture of stars from both disks
and from the halo. In very rough numbers we find that ten out of a
hundred stars are thick disk stars and the rest are thin disk and that
one out of a thousand solar neighbourhood stars are halo stars (Buser
2000).  If we want to study the elemental abundances in stars that
we believe belong to the thick disk we need to decide how to select
appropriate targets. There are essentially two ways to do that:
{\bf Position} -- sufficiently high above the plane that the star is
  more likely a thick disk than a thin disk star, or,
{\bf Kinematics} -- various kinematic criteria may be formulated to
  distinguish the disks from one another.

  Different kinematic criteria have been used by different
  authors. The following three examples highlight some of the
  differences and similarities. It is interesting to note, however,
  that the resulting abundance trends essentially show the same
  results. 

  {\bf Bensby et al. 2003 \& 2005} All their stars are from the
  Hipparcos Catalogue and have radial velocities as well as
  Str\"omgren photometry available in the literature. For these stars
  a kinematic selection was done in the following way: assume the
  velocity components have Gaussian distributions unique to each
  population (i.e. halo, thick disk, thin disk); allow for the
  different asymmetric drifts; calculate the probability that each star
   belongs to the halo, thick disk, and thin disk,
  respectively; for thick disk component they selected stars that were
  more likely to be thick disk than thin disk, and vice verse for the
  thin disk. It turns out that the selection is not very sensitive to
  the local normalisations of the number densities for the disks
  (Bensby et al.\,2005). This also shows, as can be expected from the
  procedure itself, that two fairly extreme samples are selected.

  {\bf Gratton et al. 2003} Using accurate parallaxes from the
  Hipparcos catalogue and radial velocities from the literature were
  used to calculate the orbital parameters and space velocities for
  the stars. The stars were then subdivided into three categories: An
  inner rotating population; a second population containing
  non-rotating and rotating stars; the third category finally contains
  the thin disk. This last category is confined to the galactic plane
  as defined by the orbital parameters of the stars (i.e. maximum
  height reached above the plane and eccentricity). The second
  category is identified as the halo and the first includes part of
  what is in the two other studies called halo and all of their thick
  disk. This population is referred to as the dissipative component as
  they are not able, in their kinematic as well as abundance data, to
  find any discontinuity between what is generally called the halo and
  the thick disk.

{\bf Reddy et al. 2003 \& 2006} All stars are from the Hipparcos
Catalogue and have radial velocities available in the literature. A cut
for distances, at 150 pc, was imposed to avoid problems with
reddening. An initial selection of stars belonging to the thin and
thick disks, respectively, is done by imposing cuts in $V_{\rm LSR}$
and $W_{\rm LSR}$. These selections are also ``verified'' by computing
probabilities akin to those in e.g. Bensby et al.\,(2003).

These studies also impose criteria such that only stars within
a fairly narrow range of effective temperature and $\log g$ are
selected. Hence, it become possible to 1) do a differential study and
2) obtain ages for the stars based on their position in the
HR-diagram.

Additionally, Klaus Fuhrmann has in a series of papers (Fuhrmann 1998;
2000 unpublished; and 2004) investigated a sample containing all mid-F
to early K dwarf stars within 25 pc and with $M_V=6$ and north of
declination $\delta=15^\circ$. As such a sample is mainly made up of
thin disk stars he has added stars at larger distances that are
assumed representative of the thick disk and halo. However, the basic
criteria to assign a star to either thin or thick disk have evolved
between the papers. In the first paper the chemical signatures
(i.e. Mg abundances) were the major criteria whilst in the second
paper age is envisioned as the criterion that will distinguish a star
as either thin or thick disk. It also turns out that these
assignments do agree with kinematic classifications based on
e.g. $V_{\rm LSR}$ and total velocity.

However, it appears to be a more robust and straightforward method to
first identify stars according to reproducible kinematic criteria and
then study their abundances and ages as we do not apriory have
knowledge about what the thick disk is but want to find out.

\section{The abundance trends in the thick and thin disks}

\begin{figure}
\begin{center}
\includegraphics[width=4in]{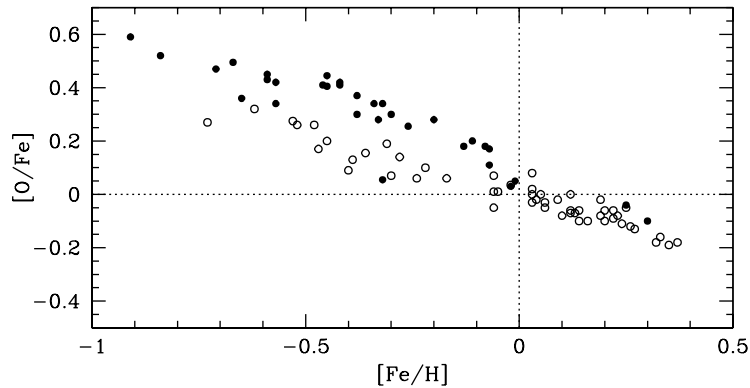}
\end{center}
\caption{[O/Fe] vs.\ [Fe/H] for stars with kinematics typical of the
  thick disk (marked with $\bullet$) and stars with kinematics typical
  of the thin disk (marked with $\circ$). The data are taken from
  Bensby et al. (2004a \& 2005).}\label{fig.oxygen}
\end{figure}

Recent studies of the elemental abundance trends in the thin and thick
disks include the following differential studies: Fuhrmann (1998,
2004), Chen et al. (2000), Mashonkina et al (2003), Gratton et
al. (2003), Bensby et al (2003, 2004a, 2005), Bensby \& Feltzing
(2006), Feltzing et al. (2006), and Mishenina et al. (2004). These are
complemented by studies that have focused on only one of the disks. The
two most important studies of only thick disk stars are Prochaska et
al. (2000) and Reddy et al. (2006). For the thin disk Reddy et
al. (2003) and Allende Prieto et al. (2004) are of particular
interest.

The main findings from these studies may be summarized as follows:

\begin{enumerate}

\item at a given [Fe/H] the stars with kinematics typical of the thick
  disk are more enhanced in $\alpha$-elements than the stars with
  kinematics typical of the thin disk (e.g. Bensby et al. 2003 \&
  2005; Fuhrmann 1998 \& 2004; Gratton et al. 2003)

\item other elements also show differences for the two disks, e.g. Ba,
  Al, Eu, and Mn (Mashonkina et al. 2003; Bensby et al. 2005; Feltzing et
  al. 2006; Feltzing 2006)

\item the elemental abundance trends for the kinematically selected
  samples are tight (Bensby et al. 2004a; Reddy et al. 2006)

\item studies that follow stars with kinematics typical of the thick
  disk up till solar metallicities note a downward trend in
  e.g. [O/Fe] as a function of [Fe/H]. This is most easily interpreted
  as a contribution from SN\,Ia to the chemical enrichment
  (Fig.\ref{fig.oxygen} and Bensby et al. 2003)

\end{enumerate}

An essential part of the studies cited in the list above is that they
all employ a differential method. That is, in the study both stars
with kinematics typical of the thin disk as well as stars with
kinematics typical of the thick disk are included. Furthermore, the
stars only span a narrow range in effective temperature and $\log
g$. This means that, to first order, any modelling errors in the
abundance determination cancel. It is important to note that when data
from different studies are combined often the distinct trends are
blurred as it is very difficult to fully put different studies on the
same base-line as concerns the derived abundances.

It is interesting and important to note that although different
studies apply different kinematic selection criteria the results are
robust and remain the same. This implies that the currently used
criteria are selecting essentially the same stellar populations. The
important fact we have learnt in the last decade is that stars that
occupy the velocity space associated with the thick disk show
elemental abundance trends that are distinct from the trends traced by
stars with kinematics typical of the thin disk.

\subsection{Vertical structures}

\begin{figure}
\begin{center}
\includegraphics[width=3.75in]{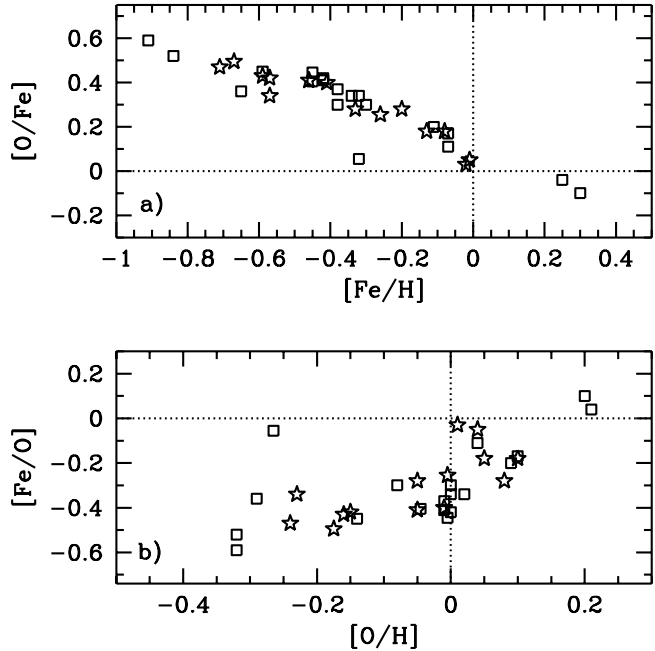}
\end{center}
\caption{[O/Fe] vs.\ [Fe/H] and [Fe/O] vs.\ [O/Fe] for all stars with
  thick disk kinematics in Bensby et al. (2003, 2004a \& 2005). The
  stars have been divided according to how far they reach above the
  galactic plane. Stars with $Z_{\rm max} > 500$ pc are marked by open
  squares and $Z_{\rm max} \leq 500$ pc are marked by open
  stars. Compare also Fig.\,\ref{fig.height2} which shows the $W$
  velocities and estimated distances from the galactic plane for these
  stars as well as for the stars with kinematics typical of the thin
  disk from the same studies.}\label{fig.height}
\end{figure}

\begin{figure}
\begin{center}
\includegraphics[width=3.75in]{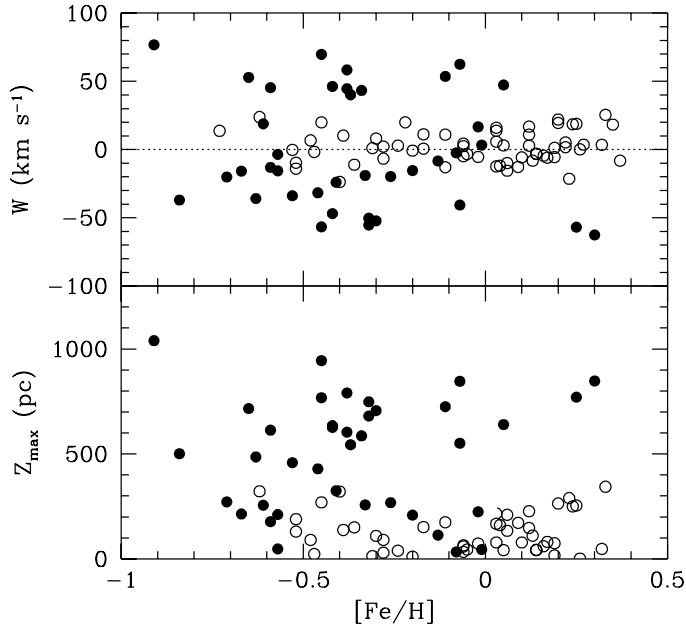}
\end{center}
\caption{For all stars in Bensby et al. (2003 \& 2005) we show in the
  top panel the $W$ velocity as a function of [Fe/H] and in the bottom
  panel an estimate of the $Z_{\rm max}$ these stars will reach based
  on their measured $W$ velocities. For details see Bensby et
  al. (2005). Stars that have kinematics like the thick disk are shown
  as $\bullet$ and stars with kinematics like the thin disk are marked
  with $\circ$. For the thick disk $\sigma_{\rm W} \sim 35$ km
  s$^{-1}$ }\label{fig.height2}
\end{figure}

\begin{figure}
\begin{center}
\includegraphics[width=5.5in]{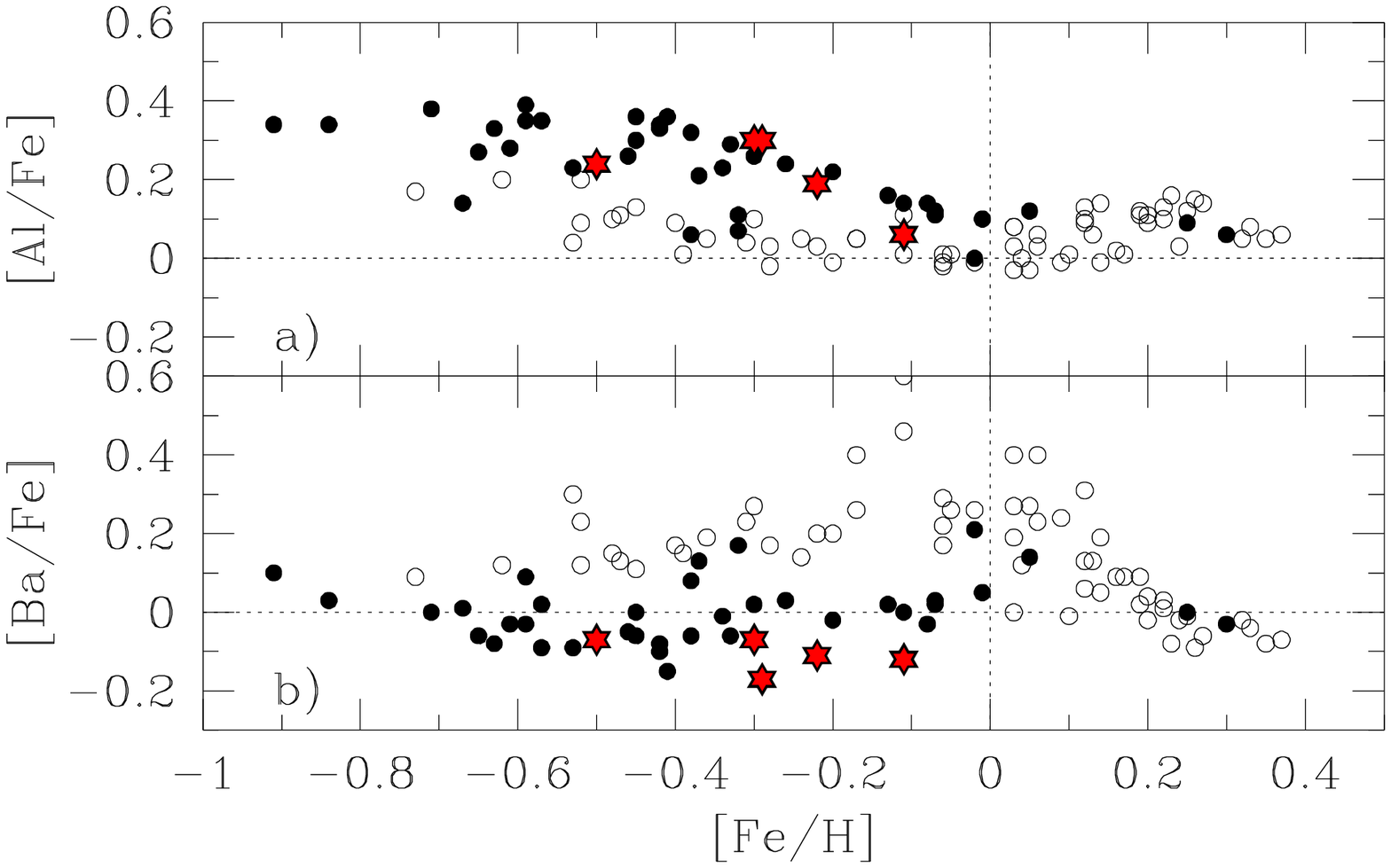}
\end{center}
\caption{[Al/Fe] and [Ba/Fe] vs.\ [Fe/H] for three stellar samples.
  $\bullet$ marks local solar neighbourhood stars with kinematics
  typical of the thick disk and $\circ$ mark stars with kinematic
  typical of the thin disk. Both samples are from Bensby et al. (2003
  \& 2005). Red/grey stars mark the five dwarf stars at the South
  Galactic Pole for which we have obtained spectra (Feltzing et al. in
  prep). The stars are on average $\sim$ 1 kpc away from the
  Sun. }\label{fig.bafeheight}
\end{figure}

Changes in the properties of the stellar populations as a function of
height above the galactic plane are important clues to the formation
of the galactic disk system. A slow, monolithic collapse would for
example result in clear trends such that the mean metallicity would
increase with decreasing distance from the galactic plane. If instead
the thick disk was formed from an originally think disk that was later
puffed up in a merger event we should see no such trends. In that case
we would also expect the abundance trends in the thick disk to be the
same at all heights.

Gilmore et al.\,(1995) studied the metallicity distribution function
at 1.5 and 2 kpc above the galactic plane. They found no differences
between the two distributions.  Davidge (2005) and Mould (2005) also
find that there is no appreciable gradient in the colours of the
stellar populations as a function of the height above the galactic
plane in nearby spiral galaxies.  This is consistent with the result
found for the Milky Way by Gilmore et al. (1995).

In Fig.\ref{fig.height} we have divided the stars with kinematics
typical of the thick disk into two samples based on how far their
$W_{\rm LSR}$ velocities will take them above the galactic plane
(Bensby et al. 2005). The two samples show exactly the same abundance
trends. These findings appear to exclude a monolithic collapse for
the formation of the thick disk and favour a puffing-up scenario.
Tentative results from an abundance study of 5 dwarf stars situated
above the galactic plane at $\sim 1$ kpc show that the elemental
abundance trends for these stars are the same as for the local,
kinematically selected thick disk stars (Feltzing et al. 2006, to be
submitted to A\&A). Figure\,\ref{fig.bafeheight} shows the results for
Ba and Al. The [Ba/Fe] vs. [Fe/H] trend for the local, kinematically
selected thick disk stars is well separated from that of the thin disk
stars at [Fe/H]$\sim 0$. The five ``in situ'' dwarf stars clearly
follow the local thick disk trend rather than the local thin disk
trend. Also for Al we see a clear separation of the two trends and the
stars at $\sim 1$ kpc show the same trend as the stars with kinematics
typical of the thick disk.

\subsection{How metal-rich can the thick disk be?}

An interesting and unanswered questions is: How metal-rich are the most
metal-rich stars in the thick disk? When selecting stars with
kinematics typical of the thick disk we do find stars with typical
thick disk kinematics at up to [Fe/H]=0 and even a few stars above
solar metallicity (e.g. Bensby et al. 2005). That such stars really
belong to the thick disk has been questioned. Rather it could be
argued that they belong to the tail of the velocity distribution of
the thin disk (see e.g. discussion in Mishenina et al. 2004).
Figure\,\ref{fig.height2} shows the relevant kinematic data and
estimated distance reached above the galactic plane for all stars in
Bensby et al. (2003 \& 2005)

However, tentative results for stars $\sim$ 2 kpc above the galactic
disk show that a large portion of such stars also have solar
metallicities (Arnadottir, Feltzing et al. in prep.). If the number of
metal-rich stars is compatible with the number of expected thin disk
stars at these heights remains to be confirmed. Also, when inspecting
large kinematic samples (e.g. the sample collected in Bensby et
al. 2004b) it is clear that stars with $|W_{\rm LSR}| > 35$ km
s$^{-1}$ (the velocity dispersion for the thick disk) are frequent
also at solar metallicities. However, also here detailed modelling is
needed to draw any firm conclusions. For now, we feel that it might be
most productive to keep an open mind in this particular issue and
investigate the metal-rich thick disk further.

\section{Ages}

For dwarf stars that have evolved of the main sequence it is possible,
even though difficult, to derive their ages (J\o rgensen \& Lindegren
2005).  Also for the studies of the age structure(s) in the disks it
is important to do differential studies, i.e that all stellar
parameters are derived in the same way and that all ages are derived
using the same isochrones. In this way modelling errors will cancel
(at least to first order) and we can say with confidence which stars
are the older ones and find out if there are any trends such that
e.g. more metal-rich stars are younger. It is also important to take
$\alpha$-enhancement into account as that tends to make a star younger
(see e.g. Kim et al. 2002; Yi et al. 2001).

So far, in all abundance studies of the thin and the thick disk it is
found that the stars with kinematics typical of the thick disk are
older than the stars with kinematics typical of the thin disk
(e.g. Fuhrmann 2004; Bensby et al. 2005).  Recent studies of resolved
stellar populations in nearby spiral galaxies show that their thick
disks also appear to be all old, Davidge (2005) and Mould (2005).

If there is hiatus in the star formation such that there is an age gap
between the thin and the thick disk is debated (compare e.g. Gratton
et al. 2003 and Bensby et al. 2004b \& 2005). Here exact selection
criteria might play a r\^ole.

More comprehensive studies of the age properties of the thick disk
component have been done by Bensby et al. (2004b), Schuster et
al. (2006), and Haywood (2006). In all of these studies it is found
that there appears to be an age-metallicity relation present in the
stellar population with kinematics typical of the thick disk. The
change in ages might be as large as 3--4 Gyr (Bensby et
al. 2004b). Thus, it does appear that the star formation in this stellar
population has been extended over time.

\section{Discussion}

The two most important features of the disk that any model must be
able to reproduce is the fact that tight and different trends are
observed for elemental abundances (e.g. oxygen) for stars with
kinematics typical of the thin and the thick disks, respectively, and
that there are no evidence that either the elemental abundance trends
or the metallicity distribution functions in the thick disk vary with
height above the galactic plane (Bensby et al. 2005; Gilmore et al. 1995).

Additional constraints are provided by the stellar ages. It appears
that the stellar population in the thick disk is older than that in
the thin disk (e.g. Fuhrmann 2004; Bensby et al. 2004b). Also, there
is evidence for an age-metallicity relation in the thick disk
(Schuster et al. 2006; Haywood 2006; Bensby et al. 2004b).

Several scenarios for the formation of thick disks in spiral galaxies
have been suggested. A comprehensive, and still valid, summary can be
found in Gilmore et al.\,(1989). Earlier work focused on various
versions of monolithic collapse. Although the observational evidence
is still somewhat meagre it does appear that the fact that to date no
vertical gradients have been observed invalidates this
scenario. Recent efforts have focused on $\Lambda$CDM realisations of
the formation of Milky Way like galaxies (e.g. Abadi et al. 2003,
Brooks et al. 2004 \& 2005). These models suggest that the thick disk
in the solar neighbourhood could be made up of a single accreted dwarf
galaxy in which the stars had formed prior to the merger but that
other parts of the thick disk, e.g. closer to the Bulge, would
originate in other dwarf galaxies. If this is correct we should expect
to see rather different abundance patterns in different parts of the
Milky Way thick disk as there is no reason to believe that the merged
galaxies would have the same potential wells and hence produce
identical abundance patterns. In general, if thick disk stars are
accreted to the Milky Way they must come (at least the ones close to
the Sun) from the same object as it would be hard to imagine how to
create the tight abundance trends that we observe in the local thick
disk. Other models, e.g. Kroupa (2002), suggest that the Milky Way has
gone through phases of enhanced star formation due to close encounters
between the Milky Way and a passing satellite galaxy. The star
formation rate would not only be enhanced in these models but gas
would also be stirred and exited to higher latitudes in the Milky Way,
thus creating the thick disk stars ``in situ''.

A widely advocated scenario is that what we today see as the thick disk
originally was a thin disk that was heated by a minor merger between
the Milky Way and a satellite galaxy. However, it currently appears
that the thick disk is all old. This would mean that the last merger
happened long ago (3--4 Gyr if we should believe the current age
estimates). This in turn would not work well with the $\Lambda$CDM
models. Another constraint is that the stars that make up
our local thick disk must have formed in a fairly deep potential well
as the mean metallicity of the thick disk is high (Wyse 1995). This in
combination with the clear enhancement of the elemental abundances for
the $\alpha$ elements which indicates that SN\,II dominated the
chemical enrichment indicates that the if the thick disk stars did
form in a satellite it must be large (Wyse 2006).

\section{Summary}

Our current knowledge about the thin and the thick disks can be
summarised as follows.  The summary has been divided into three
categories:

\bigskip
\noindent
\textit{Controversial findings/claims:}

\begin{itemize}
\item The thick disk extends to [Fe/H]=0
\item There is an age-metallicity relation present in the thick disk
\end{itemize}

\medskip
\noindent
\textit{The following are items that are fairly well agreed upon:}

\begin{itemize}
\item The thick disk shows evidence for extended star formation

\item No changes in abundance trends and/or metallicity distribution
  functions as a function of height above the galactic plane have been
  found (yet)

\end{itemize}

\medskip
\noindent
\textit{Commonly acknowledged as well established are:}

\begin{itemize}
\item Abundance trends for kinematically selected samples differ

\item The elemental abundance trends in the kinematically selected
	samples are very tight

\item Stars with kinematics typical of the thick disk are enhanced in
  $\alpha$-elements as compared to the stars with kinematics typical
  of the thin disk (at a given [Fe/H])

\item The solar neighbourhood thick disk stars that have been studied are all old

\item To date all stars with kinematics typical of the thick
disk that have been studied with high resolution spectroscopy
appear to be older than those with kinematics typical of the
thin disk
\end{itemize}

\bigskip

The real challenge for models of galaxy formation is to explain the
tight elemental abundance trends found for kinematically selected
populations of disk stars, i.e. the thin and the thick disk.  There is
also a need to be able to accommodate the observed age constraints,
i.e. the stars in the thick disk are older than those in the thin disk
and perhaps that there is an age-metallicity relation present in the
thick disk.

A wealth of information about the stars in the Milky Way disks have
been collected so far. However, as the discussion above suggest we are
still quite far away from our goal -- to understand how the two disks
formed and evolved. The current observational evidence does not even
appear to be quite strong enough to distinguish between some of the
major formation scenarios that have been proposed so far. 

Future progress will depend on access to large surveys (see
e.g. various contributions to the Joint Discussion 13 held at the IAU
General Assembly in Prague 2006
{\tt http://clavius.as.arizona.edu/vo/jd13/}) but it will also, in equal,
measures depend on the quality of the data (i.e. high resolution,
high signal-to-noise), and the treatment of the data (i.e. the
modelling of stellar atmospheres and line formation). This includes
obtaining improved atomic data, 3D modelling of stellar atmospheres,
and NLTE treatment of line formation.

\begin{acknowledgements}
SF is a Royal Swedish Academy of Sciences Research Fellow supported
by a grant from the Knut and Alice Wallenberg Foundation.
\end{acknowledgements}

\end{document}